\documentclass[aps,prb,twocolumn]{revtex4}
\usepackage{epsfig}
\begin{document} 
\title{Ferromagnetism in the {\mbox{\boldmath $t$-$t'$}} Hubbard model: 
interplay of lattice, band dispersion, and interaction effects 
studied within a Goldstone-mode preserving scheme}
\author{Sudhakar Pandey$^1$}
\email{spandey@iitk.ac.in} 
\author{Avinash Singh$^{1,2}$}
%\email{avinas@iitk.ac.in} 
\affiliation{$^1$Department of Physics, Indian Institute of Technology Kanpur - 208016}
\affiliation{$^2$Max-Planck-Institut f\"{u}r Physik Komplexer Systeme, N\"{o}thnitzer str. 38, D-01187 Dresden}
\begin{abstract}
Ferromagnetism in the Hubbard model is investigated on sc, bcc, and fcc lattices 
using a systematic inverse-degeneracy ($1/{\cal N}$) expansion 
which incorporates self-energy and vertex corrections 
such that spin-rotation symmetry and the Goldstone mode are explicitly preserved. 
First-order quantum corrections to magnon energies are evaluated for several cases,
providing a comprehensive picture of the 
interplay of lattice, band dispersion, and interaction effects
on the stability of the ferromagnetic state 
with respect to both long- and short-wavelength fluctuations.
Our results support the belief that ferromagnetism is a 
generic feature of the Hubbard model at intermediate and strong coupling
provided the DOS is sufficiently asymmetric and strongly peaked near band edge,
as for fcc lattice with finite $t'$. 
For short-wavelength modes, 
behavior of a characteristic energy scale $\omega^* \sim T_c$ (magnon-DOS-peak energy)
is in excellent agreement with the $T_c$ vs. $n$ behavior within DMFT,
both with respect to the stable range of densities ($0.20 < n < 0.85$) as well as 
the optimal density $n=0.65$.
However, our finding of vanishing spin stiffness near optimal density
highlights the role of long-wavelength fluctuations in further 
reducing the stable range of densities.
\end{abstract}
\maketitle
\section{INTRODUCTION}
Band ferromagnetism involves a characteristic competition between 
decrease in interaction energy and increase in kinetic (band) energy 
resulting from unequal spin densities in the ferromagnetic state. 
The Stoner criterion $UN(\epsilon_{\rm F}) = 1$,
which signals the onset of local moments and highlights the strong-coupling character, 
represents the simplest (Hartree-Fock) expression of this competition, 
with the corresponding $T_c ^{\rm HF}$ representing a moment-melting temperature. 
The physical $T_c$, on the other hand, represents a spin-disordering temperature,
and is typically (even at RPA level) much lower than $T_c ^{\rm HF}$ 
due to strong transverse spin fluctuations 
resulting from the low-energy scale of collective excitations (magnons), 
which also significantly increase beyond the Stoner value $1/N(\epsilon_{\rm F})$ 
the critical interaction strength $U_c$ required for long-range magnetic order. 
The requirement of self-consistently incorporating,
in the non-perturbative strong-coupling regime, 
spin and charge fluctuations in the magnon description,
while maintaining the essential spin-rotation symmetry,
summarizes, from the broken-symmetry side, 
the challenging nature of band ferromagnetism. 

Rigorous results for ferromagnetism in 
the Hubbard model have been obtained only in a few special cases. 
These include the Nagaoka theorem for a single hole in a half-filled band
in the singular $U \rightarrow \infty$ limit,\cite{Nagaoka}
%The extension of this result, in general, 
%is not clear for finite $U$ and finite particle density. 
Lieb ferrimagnetism\cite{Lieb} at any $U > 0$ for half-filled bipartite lattices 
having asymmetry in the number of sites per sublattice, 
flat-band ferromagnetism\cite{Mielke1,Mielke2,Tasaki1,Mielke-Tasaki} 
in specially decorated lattices having dispersionless spectrum and special band filling 
and its extension to nearly flat-band systems.\cite{Tasaki2,Tasaki3,Mielke3,Tanaka-Ueda,Tasaki4} 

%Our understanding of metallic ferromagnetism has been further improved 
%with the newly developed many-body techniques such as 
%dynamical mean-field theory (DMFT),\cite{Metzner-Vollhardt,GKKR-RMP,Uhrig} 
%which becomes exact in the limit of infinite dimensions ($d=\infty$), 
%and density matrix renormalization group\cite{White}(DMRG), 
%which yields precise results in one dimension.\cite{Daul-Noack}($d=1$) 
%However, the situation in case of the realistic interests ($d=2, 3$) 
%is still far from clear.

Recent reports of ferromagnetism in the single-band Hubbard model on fcc lattice 
with next-nearest-neighbour hopping $t'$ in the intermediate coupling regime, 
and over a wide range of electronic densities,\cite{Ulmke,Arita,Rumch}  
using a variety of techniques such as dynamical mean field theory (DMFT),
fluctuation-exchange (FLEX) approximation,
and modified random phase approximation (MRPA), 
have renewed interest in the classic problem of metallic ferromagnetism,
and have highlighted the importance of lattice structure
and band dispersion in the origin of ferromagnetism. 
There is now substantial evidence that non-bipartite lattices such as the fcc lattice, 
having asymmetric density of states (DOS), 
with Fermi energy located in the region of large spectral weight near the band-edge, 
are favorable for ferromagnetism.\cite{Ulmke,Arita,Rumch,Shastry,Muller-physica,Vollhardt-jltp,Uhrig,Nolting-solid,Muller-prb,Vollhardt-prb,Vollhardt-cond} 
The asymmetry in DOS is further enhanced by the next-nearest-neighbor hopping term ($t'$),
providing optimal condition for ferromagnetism 
in the weak and intermediate coupling regimes.\cite{Ulmke,Arita}  
Similar enhancement of ferromagnetism in the intermediate coupling regime
due to $t'$-induced DOS asymmetry has been reported recently
for the one-dimensional lattice\cite{Daul-Noack} 
and for the square lattice.\cite{Hlubina1,Markus,Hlubina2,Arrachea,Irkhin,Honerkamp,Katanin,Hankevych,Taniguchi} 
Bipartite lattices with symmetric DOS such as sc and bcc lattices are,
on the other hand, controversial candidates in which ferromagnetism exists 
only in the strong-coupling regime, 
if at all.\cite{Shastry,Vollhardt-jltp,Nolting-solid,Obermeier-largeU} 

The recent investigations of existence of ferromagnetism on the fcc lattice 
have also highlighted the subtle dependence on the 
distribution of spectral weight in the DOS induced by $t'$. 
Ferromagnetism was found to be absent for $t'=0$
in both DMFT\cite{Ulmke} and FLEX\cite{Arita} studies,
although at sufficiently large interaction strength $(U/W \geq 2.5)$
a stable ferromagnetic state was obtained in the MRPA study.\cite{Rumch} 
While a stable ferromagnetic state was obtained in a wide density range 
for $t'=t/4$ in the DMFT study 
for $U$ above a critical value $U_c \sim W$ in the intermediate coupling regime,
in the FLEX study for $t'=t/2$ in the weak coupling limit ($U << W$), 
ferromagnetic fluctuations were found to be strongest at $n=0.2$. 
Also, the optimal density as well as the range of densities over which  
ferromagnetism is stabilized differ substantially in these investigations.

However, there are certain limitations in the approaches (DMFT, FLEX, MRPA) discussed above. 
Incorporating only the local (Ising) spin excitations, 
DMFT ignores long-wavelength spin fluctuations and the ${\bf k}$-dependence of self energy. 
FLEX incorporates ${\bf k},\omega$-dependence of self energy, 
but ignores vertex corrections of the same order, 
thereby breaking the spin-rotation symmetry. 
Both DMFT and FLEX are therefore not in accordance with the Mermin-Wagner theorem, which 
rules out long-range magnetic order in one and two dimensions at any finite temperature. 
While an improved form of self-energy correction has been incorporated in MRPA, 
the momentum and frequency dependence of vertex corrections have been ignored. 

In the context of self-energy corrections in a band ferromagnet, 
the importance of vertex corrections in restoring the spin-rotation symmetry 
and Goldstone mode has been long recognized at a formal level in terms of a Ward identity 
explicitly connecting vertex corrections to self-energy corrections.\cite{Hertz} 
Although the spin stiffness was shown to be reduced from its RPA value, 
spin-wave excitations for arbitrary wave-vector were not considered, 
and also contribution of Stoner excitations were ignored, 
which can have significantly adverse effect on the stability of the ferromagnetic state,
as shown in Sections III and IV. 

The objective of the present paper is to quantitatively investigate 
correlation effects in a band ferromagnet within a spin-rotationally-symmetric scheme 
compatible with the Mermin-Wagner theorem.
Such a scheme is provided by the inverse-degeneracy $(1/{\cal N})$ expansion
within the generalized ${\cal N}$-orbital Hubbard model,\cite{As-quantum} 
where quantum corrections beyond the RPA can be systematically incorporated
while explicitly preserving spin-rotation symmetry and hence the Goldstone mode 
order-by-order. This approach has been applied earlier to examine quantum corrections 
in the antiferromagnetic state of the Hubbard model,\cite{As-quantum}
where it was shown that in the strong-coupling limit 
the $1/{\cal N}$ expansion becomes equivalent to the $1/2S$ expansion for
the spin-$S$ quantum Heisenberg antiferromagnet.  
Recently, first-order quantum corrections in the ferromagnetic state have been 
obtained diagrammatically by including all O$(1/{\cal N})$ self-energy and vertex 
corrections.\cite{As-goldstone} 
Within this approach, the $q^2$ Goldstone-mode spectrum for the ferromagnet 
results in diverging transverse spin fluctuations 
and hence absence of long-range magnetic order
at any finite temperature in one and two dimensions,
exponentially large spin-correlation length in two dimensions,
and the usual $T^{3/2}$ Bloch-law behavior of magnetization 
at low temperature in three dimensions 
due to thermal excitation of low-energy, long-wavelegnth modes. 
This spin rotationally-symmetric approach is thus especially suitable for 
low-dimensional, low-temperature, and long-wavelength studies,
and provides a complimentary approach to DMFT.

In this paper we examine the first-order $(1/{\cal N})$ quantum corrections
to the collective spin-wave excitations (magnons) in the saturated 
ferromagnetic ground state of the $t$-$t'$ Hubbard model.
Our investigations for different lattices (sc, bcc, fcc), 
next-nearest-neighbor hoppings ($t')$, interaction strengths ($U$), and wave-vectors ($q$)
provide a comprehensive picture of the interplay of lattice, band dispersion, 
and interaction effects on the stability of the ferromagnetic state 
with respect to both long- and short-wavelength fluctuation modes.
%Our specific choice of saturated ground state, where all 
%the minority-spin states are pushed above the Fermi energy, results in the relative 
%simplifications of  the analytical and numerical analysis. 
The organization of this paper is as follows. 
In Sec. II we briefly review the spin-rotationally-symmetric approach 
to the transverse spin fluctuation propagator.
Section III A provides a review of the RPA-level (classical) contributions
(delocalization and exchange) to the spin stiffness.
The classical exchange contribution is shown to be relevant in determining the 
magnitude of the O$(1/{\cal N})$ quantum corrections, which are derived 
and examined for different cases in section III B, 
evaluated separately with and without the contribution of the Stoner excitations. 
Results for large-momentum, zone-boundary modes
with energy near the magnon DOS peak (dominant mode),
which is of interest as this energy scale provides a quantitative measure of $T_c$ 
within the spin-fluctuation theory, is discussed in Section IV.
Conclusions are presented in Section V.

\section{Transverse spin fluctuations}
Transverse spin fluctuations are gapless, low-energy excitations in the broken-symmetry state of magnetic systems possessing continuous spin-rotation symmetry. Therefore, at low temperatures they play an important role in diverse macroscopic properties such as existence of long-range magnetic order, temperature dependence of magnetization, transition temperature, spin correlations etc.

We consider the time-ordered, transverse spin-fluctuation propagator in the broken-symmetry state,
which describes both collective spin-wave and particle-hole Stoner excitations, 
and is given by  
\begin{equation}
\chi_{ij}^{-+}(t-t') =
i \langle \Psi_{\rm G} | T [ S_i ^- (t) S_j ^+ (t')]|\Psi_{\rm G}\rangle
\end{equation}
in terms of the fermion spin-lowering and -raising operators 
$S_i ^\mp = \Psi_i ^\dagger (\sigma^\mp/2) \Psi_i$.
The spin-fluctuation propagator in ${\bf q},\omega$ space can be expressed as 
\begin{equation}
\chi^{-+}({\bf q},\omega) = \frac{\phi({\bf q},\omega)}
{1-U\phi({\bf q},\omega)}
\end{equation}
in terms of the exact irreducible particle-hole propagator $\phi({\bf q},\omega)$,
which incorporates all self-energy and vertex corrections. 
The inverse-degeneracy expansion\cite{As-quantum,As-goldstone}
\begin{equation}
\phi = \phi^{(0)} + \left ( \frac{1}{\cal N} \right ) \phi^{(1)} 
+ \left (\frac{1}{\cal N} \right )^2 \phi^{(2)} + ...
\end{equation}
systematizes the diagrams in powers of the expansion parameter $1/{\cal N}$ which,
in analogy with $1/S$ for quantum spin systems, plays the role of $\hbar$.
As only the "classical" term $\phi^{(0)}$ survives in the $\cal N \rightarrow \infty$ limit, 
where $\phi^{(0)} ({\bf q},\omega) \equiv \chi^0({\bf q},\omega)$ represents the bare particle-hole propagator,
the usual RPA ladder series $\chi^0 ({\bf q},\omega)/1-U \chi^0 ({\bf q},\omega)$,
amounts to a classical-level description of non-interacting spin-fluctuation modes.
The first-order quantum corrections $\phi^{(1)}$,
involving self-energy and vertex corrections at O$(1/{\cal N})$ level,
have been recently obtained for a saturated ferromagnet,\cite{As-goldstone}  
and physically incorporate such effects as minority-spin spectral-weight transfer, correlation-induced exchange correction, coupling of spin and charge fluctuations, 
and magnon damping.

\begin{figure}
\begin{center}
\vspace*{-17mm}
\hspace*{-12mm}
\psfig{figure=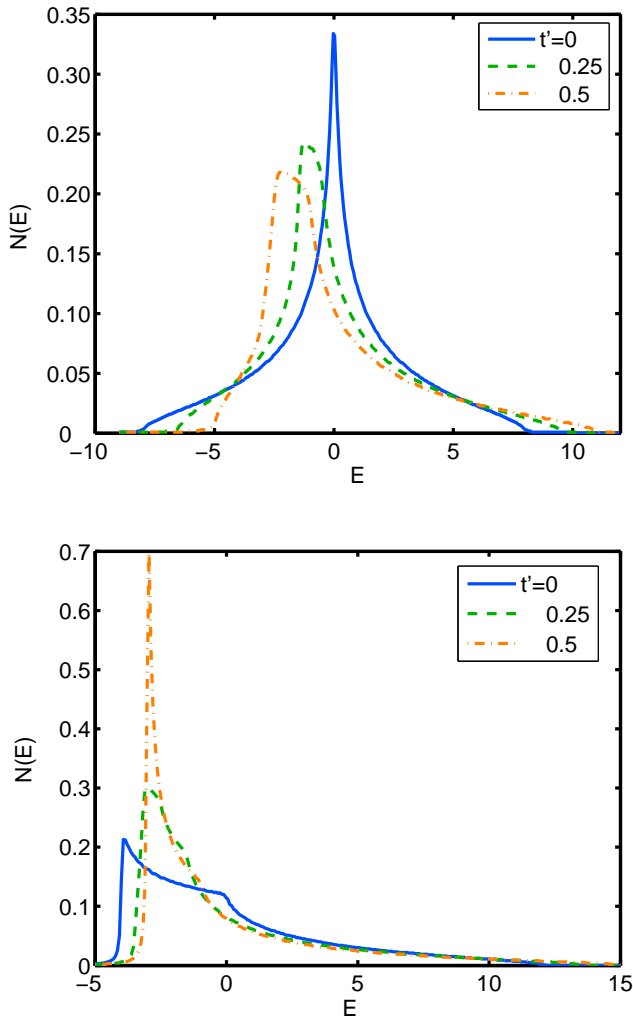,width=118mm}
\vspace*{-15mm}
\end{center}
\caption{DOS for  bcc (upper) and  fcc (lower) lattices,
showing the significant enhancement in DOS asymmetry resulting from 
redistribution of spectral weight due to the next-nearest-neighbor hopping $t'$.}
\end{figure}

As collective spin-wave excitations are represented by poles in (2), 
spin-rotation symmetry requires that $\phi = 1/U$ for $q,\omega=0$,
corresponding to the Goldstone mode.
Since the zeroth-order term $\phi^{(0)}$ already yields exactly $1/U$ for $q,\omega=0$, 
the sum of the remaining terms must exactly vanish in order to 
preserve the Goldstone mode. For this cancellation to hold for arbitrary ${\cal N}$,
each higher-order term $\phi^{(n)}$ in the expansion (3) must individually vanish,
implying that spin-rotation symmetry is preserved order-by-order,
as expected from the spin-rotationally invariant form 
$(U/{\cal N}){\bf S}_i.{\bf S}_i$ 
of the interaction term in the generalized Hubbard model.\cite{As-quantum} 
This cancellation has been shown explicitly at the O($1/{\cal N}$) level,\cite{As-goldstone}
not only for $\omega=0$ but also for all $\omega$, 
indicating no spin-wave amplitude renormalization, 
as expected for the saturated ferromagnet in which there are no quantum corrections to magnetization.

\begin{figure}
\begin{center}
\vspace*{-33mm}
\hspace*{-4mm}
\psfig{figure=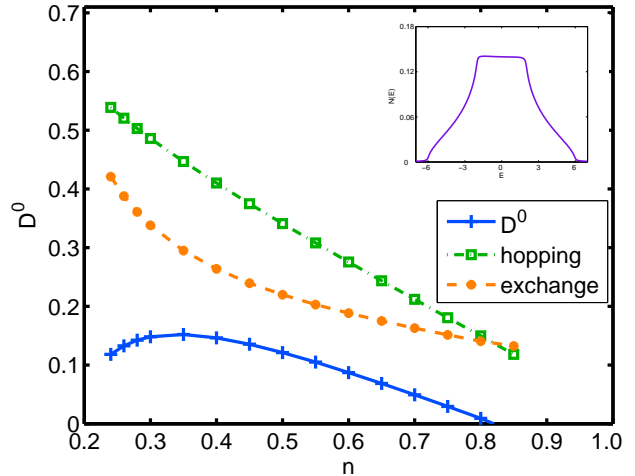,width=95mm}
\vspace*{-40mm}
\end{center}
\caption{The RPA (classical) spin stiffness $D^0$ for the 
sc lattice at $U/W =1.5$, along with the hopping and exchange contributions. 
Inset shows DOS for the sc lattice.}
\end{figure}

\section{spin stiffness}
The spin stiffness $D=\omega_{\bf q}/q^2$ in the ferromagnetic state,
defined in terms of the magnon energy $\omega_{\bf q}$ for small-$\bf q$ modes,
provides a quantitative measure of the stability of the ferromagnetic state against 
long-wavelength fluctuations, with negative $D$ signalling loss of long-range magnetic order. 
We first review the different contributions to spin stiffness 
at the RPA (classical) level 
as their behavior and interplay provide insight into the 
magnitude of the first-order quantum corrections 
and of ferromagnetic-state stability, as discussed in the next subsection.

\begin{figure}
\begin{center}
\vspace*{-17mm}
\hspace*{-3mm}
\psfig{figure=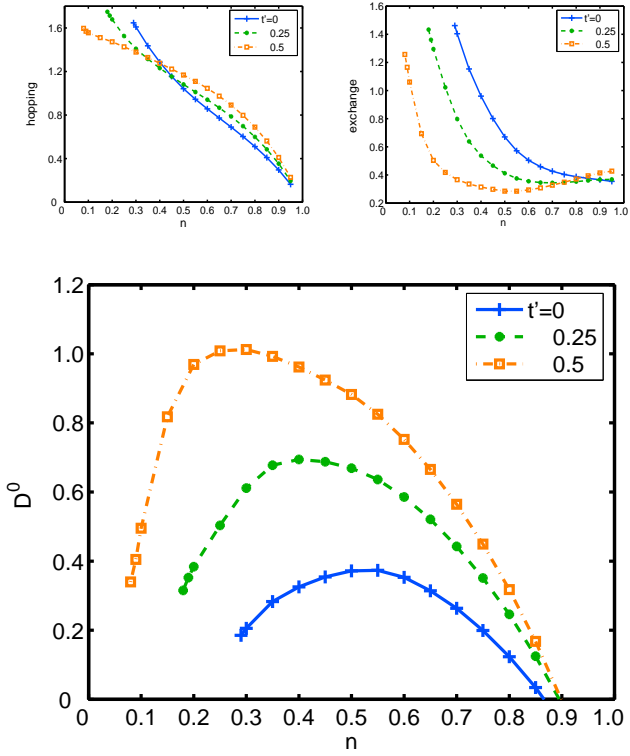,width=90mm}
\vspace*{-12mm}
\end{center}
\caption{Hopping (upper left) and exchange (upper right)
contributions to the RPA spin stiffness (lower) for the bcc lattice at $U/W=1.5$,
showing enhancement in stiffness and reduction in optimal density with increasing $t'$.}
\end{figure}

\subsection{O(1) level (RPA): interplay of hopping and exchange contributions}
The bare antiparallel-spin particle-hole propagator 
in the Hartree-Fock ground state of the saturated ferromagnet is given by 
\begin{equation}
\phi^{(0)}({\bf q},\omega) \equiv  \chi^0 ({\bf q},\omega) =
\sum_{\bf k} \frac{1}
{\epsilon_{\bf k - q}^{\downarrow +} - \epsilon_{\bf k}^{\uparrow -} + \omega -i \eta}
\; ,
\end{equation}
where normalization over number of ${\bf k}$ states is implicit in $\sum_{\bf k}$, 
$\epsilon_{\bf k}^\sigma = \epsilon_{\bf k} - \sigma \Delta$ 
are the ferromagnetic band energies, 
$2\Delta = mU$ is the exchange band splitting 
with magnetization $m$ equal to the electron density $n$,
and the superscript $+(-)$ refer to particle (hole) states above (below) 
the Fermi energy $\epsilon_{\rm F}$.
Expanding $\chi^0 ({\bf q},\omega)$ for small ${\bf q},\omega$, with 
\begin{equation}
\delta \equiv -(\epsilon_{\bf k - q } - \epsilon_{\bf k})
= {\bf q}.{\mbox{\boldmath $\nabla$}} \epsilon_{\bf k} - 
\frac{1}{2}({\bf q}.{\mbox{\boldmath $\nabla$}})^2 \epsilon_{\bf k}
\end{equation}
one obtains 
\begin{eqnarray}
& & \chi^0 ({\bf q},\omega) \nonumber \\
&=& \frac {1}{U} - \frac{1}{(2\Delta)^2} \left [
m\omega + \sum_{\bf k} \left ( 
\frac{1}{2} ({\bf q}.{\mbox{\boldmath $\nabla$}})^2 \epsilon_{\bf k} 
-
\frac{({\bf q}.{\mbox{\boldmath $\nabla$}} \epsilon_{\bf k} )^2}{2\Delta} 
\right ) \right ] \nonumber \\
&\equiv& \frac{1}{U} - {\cal A}\omega-{\cal B}^{(0)} q^2
\end{eqnarray}
%$\frac{1}{U} - {\cal A}\omega-{\cal B}{\bf q}^2$ 
which yields the RPA (classical) spin stiffness (${\cal B}^{(0)}/{\cal A}$)
\begin{equation}
D^0 = \frac{1}{d} \left [\frac{1}{2} 
\langle {\mbox{\boldmath $\nabla$}}^2 \epsilon_{\bf k} \rangle  - 
\frac{\langle ({\mbox{\boldmath $\nabla$}} \epsilon_{\bf k})^2 \rangle }{2\Delta} \right ]
\end{equation}
in $d$ dimensions. 
Here the angular bracket $\langle \; \rangle$ represents momentum summation
normalized over the number of occupied states ($\frac{1}{m} \sum_{\bf k}$). 
The two terms in (7) of order $t$ and $t^2/U$
represent hopping and exchange contributions to the spin stiffness, respectively,
corresponding to delocalization-energy loss and exchange-energy gain upon spin twisting.
As shown in section IIIB, the dominant quantum corrections involve correlation-induced
corrections to the exchange contribution. The stability of the ferromagnetic state 
therefore involves a competition between the hopping and exchange contributions,
of which the latter exhibits a subtle dependence on the band dispersion, 
distribution of spectral weight in DOS, and the underlying lattice structure.

\begin{figure}
\begin{center}
\vspace*{-12mm}
\hspace*{-2mm}
\psfig{figure=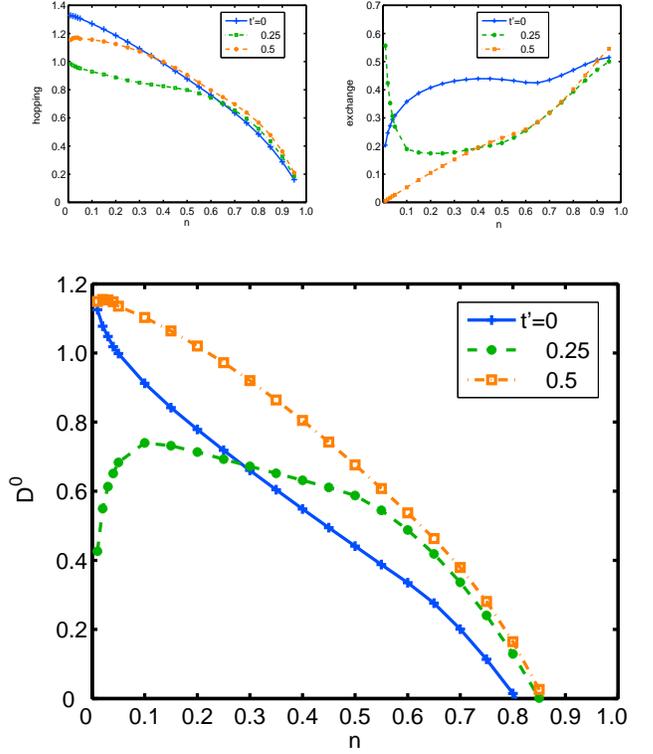,width=90mm}
\vspace*{-17mm}
\end{center}
\caption{Hopping (upper left) and exchange (upper right) contributions to the RPA spin 
stiffness (lower) for the fcc lattice at $U/W=1.0$, 
showing strong reduction in the exchange contribution 
in the low-density range for $t'=0.5$.}
\end{figure}

We have carried out investigations for sc, bcc, and fcc lattices, with energy dispersions     \begin{eqnarray}
\epsilon_{\bf k}^{\bf sc} &=& 2t\sum_\mu \cos k_\mu
\nonumber\\
\epsilon_{\bf k}^{\bf bcc} &=& 8t \cos k_x \cos k_y \cos k_z 
+2t'\sum_\mu \cos 2k_\mu
\nonumber\\
\epsilon_{\bf k}^{\bf fcc} &=& 4t\sum_{\mu < \nu} \cos k_\mu \cos k_\nu +2t'\sum_\mu \cos 2 k_\mu
\end{eqnarray}
and electronic bandwidth $W = 12t$, $16t$, and $16t + 4t'$, respectively. 
We choose energy scale such that $t=1$.
The numerical investigations have been carried out for three values of $t'$ 
(0, 0.25, and 0.5) and three values of interaction strength ($U/W$= 0.5, 1.0, and 1.5)
for each  $t'$. The motivation behind choice of these parameters is to explore the subtle dependence of quantum corrections on lattice structure, band dispersion, 
and the distribution of DOS spectral weight in the crucial intermediate-coupling regime. 
For the fcc lattice, 
the saturated-ferromagnet condition (exchange band splitting $nU$ is greater than 
Fermi energy measured from bottom of the majority-spin band)
is met in a wide range of electronic densities for all $U$ values considered.
However, for the bipartite lattices (sc and bcc with $t'=0$),
with $k^2$-like dispersion at band-edge, 
the Fermi energy ($\sim n^{2/3}$) increases more rapidly than $n$ for low densities, 
and hence comparatively larger $U$ is required to ensure the saturated-ferromagnet condition; 
therefore we have considered only $U/W=1.5$ in order to cover the maximum possible range of densities.

The limit $n\rightarrow 1$ of filled majority-spin band
is physically simple and essentially lattice independent, as seen from Figs. 2, 3, and 4.
While the hopping contribution to spin stiffness vanishes, 
the exchange contribution approaches a finite value, thereby destabilizing the ferromagnetic state. The low-density ($n \ll 1$) behavior of the exchange contribution is, however, 
highly sensitive to the distribution of spectral weight near the band bottom.
While the hopping term slowly increases to a finite value with decreasing $n$, 
the exchange contribution either increases rapidly, 
as for the sc (Fig. 2) and bcc (Fig. 3) lattices, 
and also for the fcc lattice for $t'=0.25$ (Fig. 4), 
or tends to vanish, as for the fcc lattice for $t'$ = 0 and 0.5 (Fig. 4). 
This subtle behavior of the exchange contribution is due to the  competition between the $n$-dependence of the numerator 
$\langle ({\mbox{\boldmath $\nabla$}} \epsilon_{\bf k})^2 \rangle$, 
which is highly sensitive to the distribution of spectral weight, 
and the denominator ($2\Delta = nU$) which is independent of any lattice details. 
As seen from Figs. 3 and 4,
the enhancement of spin stiffness with increasing $t'$ occurs mainly due to the significant decrease in the exchange contribution. 

As shown below, the dominant first-order quantum corrections correspond to 
correlation-induced exchange contributions, 
and the behavior of the RPA-level exchange contribution discussed above 
actually provides a qualitative idea of the magnitude of quantum corrections and therefore of the ferromagnetic-state stability.

\subsection{O($1/{\cal N}$) quantum corrections}
Expanding the net first-order quantum correction\cite{As-goldstone} 
$\phi^{(1)}$ for small ${\bf q},\omega$ in order to obtain
quantum corrections to spin stiffness, 
we find that there is a cancellation not only of 
terms of order $1/U$ and terms linear in $\omega$,
but also of terms linear in $\delta \equiv -(\epsilon_{\bf k - q } - \epsilon_{\bf k})$,
implying (from (5)) no quantum corrections to the delocalization contribution,
as expected for a ferromagnetic state.
The surviving contributions of order $q^2$ can be expressed as

\begin{widetext}
\begin{eqnarray}
\phi^{(1)}({\bf q}) &=& \frac{U^2}{(2\Delta)^4} 
\sum_{\bf Q} \int \frac{d\Omega}{2\pi i} 
\left [ \left \{ \frac{\chi^0 ({\bf Q},\Omega)}{1-U\chi^0 ({\bf Q},\Omega)} \right \}  \right .
\sum_{\bf k'} 
\frac{( {\bf q}.{\mbox{\boldmath $\nabla$}} \epsilon_{\bf k'} )^2 }
{\epsilon_{\bf k' + Q}^{\uparrow +} - \epsilon_{\bf k'}^{\uparrow -} 
- \Omega - i\eta} \nonumber \\
&-2 &  
\left \{ \frac{1}{1-U\chi^0 ({\bf Q},\Omega)} \right \}
\sum_{\bf k'} 
\frac{{\bf q}.{\mbox{\boldmath $\nabla$}} \epsilon_{\bf k'}}
{\epsilon_{\bf k' + Q}^{\uparrow +} - \epsilon_{\bf k'}^{\uparrow -} 
- \Omega- i\eta} 
\sum_{\bf k''} 
\frac{{\bf q}.{\mbox{\boldmath $\nabla$}} \epsilon_{\bf k''}}
{\epsilon_{\bf k'' - Q}^{\downarrow +} - \epsilon_{\bf k''}^{\uparrow -} 
+ \Omega- i\eta} 
\nonumber \\
&+& 
\left \{ \frac{U}{1-U\chi^0 ({\bf Q},\Omega)} \right \}
\sum_{\bf k'} 
\frac{1}
{\epsilon_{\bf k' + Q}^{\uparrow +} - \epsilon_{\bf k'}^{\uparrow -} 
- \Omega - i\eta}
\left ( 
\sum_{\bf k''} \frac{{\bf q}.{\mbox{\boldmath $\nabla$}} \epsilon_{\bf k''}}
{\epsilon_{\bf k'' - Q}^{\downarrow +} - \epsilon_{\bf k''}^{\uparrow -} 
+ \Omega- i\eta } \right )^2
\nonumber \\
&+& 
\left . 
\sum_{\bf k'} 
\frac {1}{\epsilon_{\bf k' + Q}^{\uparrow +} - \epsilon_{\bf k'}^{\uparrow -} 
- \Omega- i\eta}
\sum_{\bf k''} 
\frac{({\bf q}.{\mbox{\boldmath $\nabla$}} \epsilon_{\bf k''})^2}
{\epsilon_{\bf k'' - Q}^{\downarrow +} - \epsilon_{\bf k''}^{\uparrow -} 
+ \Omega- i\eta} \right ] \; ,
\end{eqnarray}
\end{widetext}
where we have set ${\bf q},\omega =0$ in the energy denominators 
as all terms are already explicitly second order in $q$.
All four terms in (9) represent correlation-induced exchange processes 
involving minority-spin intermediate states which are transferred down in energy. 
The spectral-weight transfer is a correlation effect 
corresponding to the possibility of a site being unoccupied by a majority-spin electron.

\begin{figure}
\begin{center}
\vspace*{-17mm}
\hspace*{-12mm}
\psfig{figure=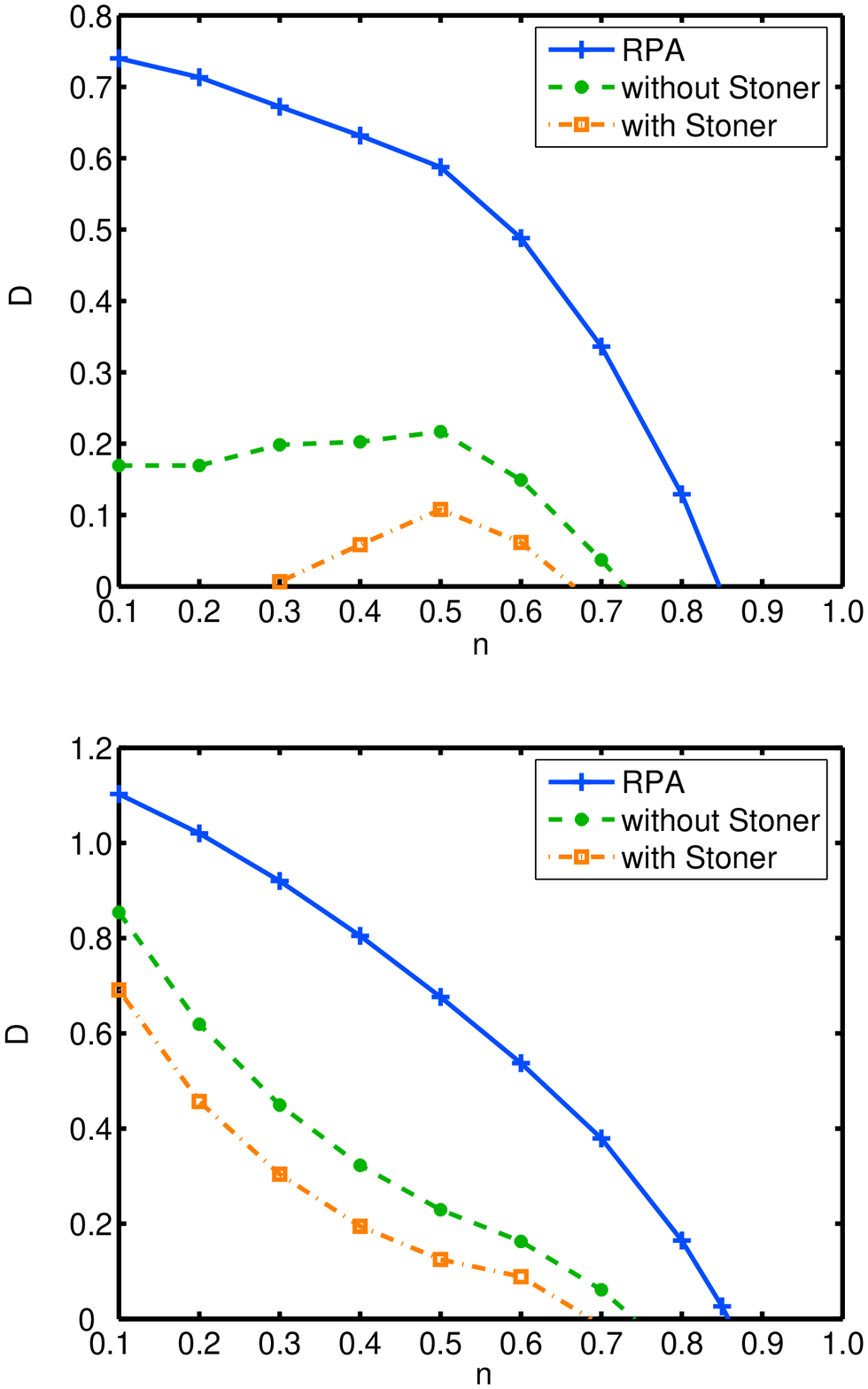,width=118mm}
\vspace*{-15mm}
\end{center}
\caption{Behavior of the corrected spin stiffness for the fcc lattice at $U/W=1.0$,
evaluated separately with and without the contribution of Stoner excitations,
shown for $t'=0.25$ (upper) and 0.5 (lower), along with the corresponding RPA results.}
\end{figure}

Now, since $\phi^{(1)}({\bf q})$ is of the form ${\cal B}^{(1)} q^2$,
the total particle-hole propagator  
$\phi =(\frac{1}{U} - {\cal A}\omega-{\cal B}^{(0)} q^2)+{\cal B}^{(1)} q^2$, 
yielding the corrected spin stiffness as 
$D=({\cal B}^{(0)}-{\cal B}^{(1)})/{\cal A}$.

In evaluating the $\Omega$-integral in Eq. (9), 
we have included the contribution of both the low-energy spin-wave excitations
as well as the high-energy Stoner excitations. 
From the spectral representation of dispersion relation,\cite{Wyld}
the $\Omega$-integral is equivalent to 
$\int_{-\infty} ^0 \frac{d\Omega}{\pi} \Im [\;]$,
and the imaginary part $\Im [\;]$ yields finite contributions 
in the negative $\Omega$ range from both collective spin-wave 
and Stoner excitations in the magnon propagator terms
$\chi^0 ({\bf Q},\Omega)/1-U\chi^0 ({\bf Q},\Omega)$ etc.,
and also from Stoner excitations in the three antiparallel-spin particle-hole 
($\sum_{\bf k''}$) terms in (9). 
If only the contribution from the low-energy spin-wave poles are included
for simplicity, the $\Omega$ integral in Eq. (9) 
can be evaluated analytically.\cite{As-goldstone}
As ${\mbox{\boldmath $\nabla$}} \epsilon_{\bf k}$ is odd in momentum, 
we find that the second and third terms in (9) give vanishingly small contributions 
due to partial cancellation, as considered earlier.\cite{As-goldstone}

%This is a good approximation when 
%collective spin-wave excitations and  Stoner excitations remain distinct in the entire MBZ. %n such cases, most of the spectral weight associated with spin-fluctuation propagator at RPA %level is contained in the spectrum of the collective spin-wave excitations.  

%However, in order to see the effects of Stoner excitations on the stability of ferromagnetic %state, particularly in  the cases when they merge with the collective spin-wave excitations %so that a significant portion of spectral weight is transferred into  the spectrum of Stoner %excitations, this integration  has  also been  evaluated numerically using the spectral %representation of dispersion relation.\cite{Wyld} 

\begin{figure}
\begin{center}
\vspace*{-17mm}
\hspace*{-10mm}
\psfig{figure=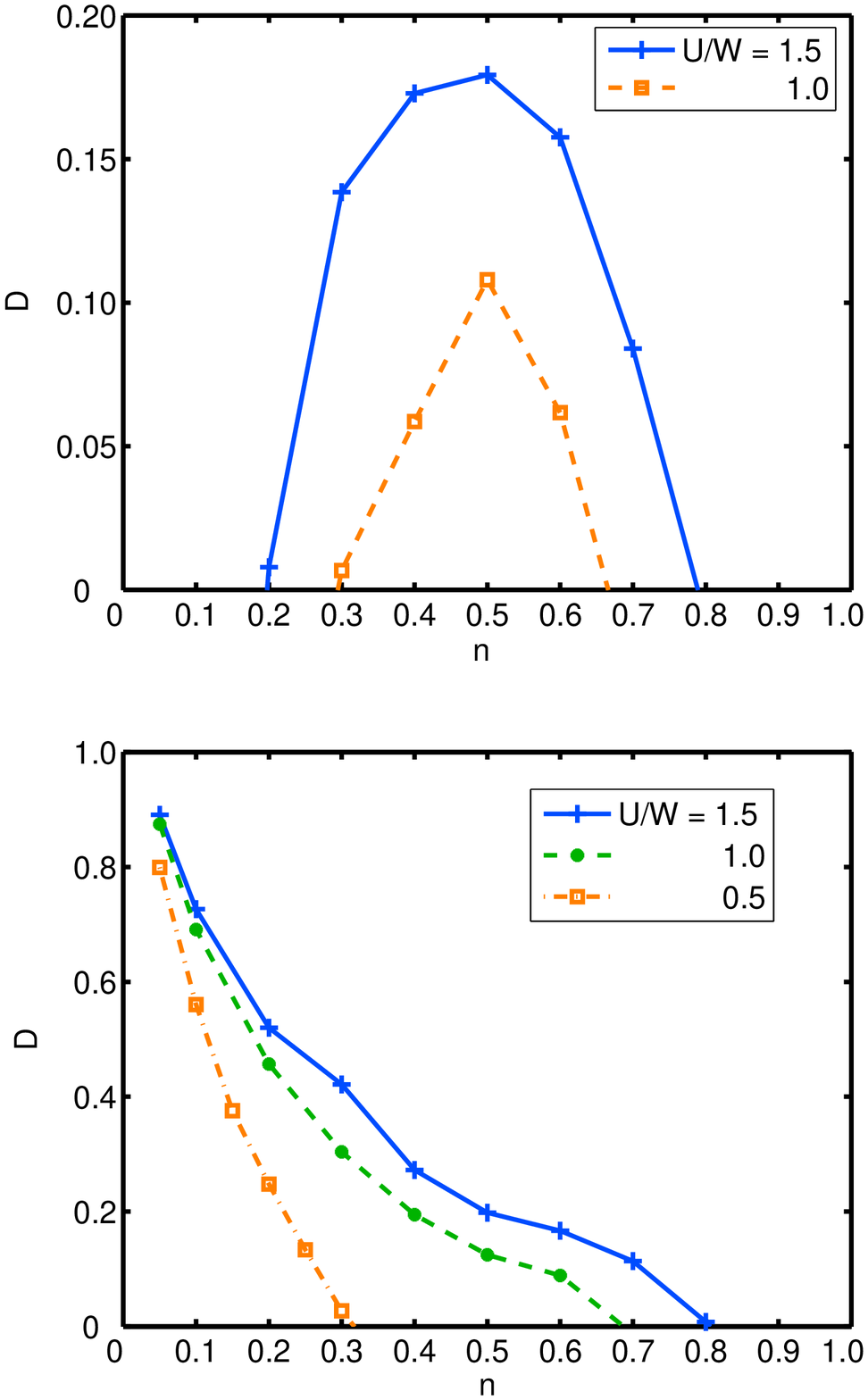,width=118mm}
\vspace*{-15mm}
\end{center}
\caption{The corrected spin stiffness for the fcc lattice shows 
strong sensitivity to $t'$, 
with stability restricted to an intermediate range of densities for $t'=0.25$ (upper) 
and to the low-density regime for $t'=0.5$ (lower).
The stiffness as well as the stable range of densities
are enhanced with increasing $U$. 
The ferromagnetic state remains unstable (negative stiffness) 
at all considered values of $U$ for $t'=0$ and also at $U/W=0.5$ for $t'=0.25$.}
\end{figure}

We find that quantum corrections are strong enough to destabilize the ferromagnetic state for the sc lattice, bcc lattice for $t'=0,0.25$, and also for the fcc lattice for $t'=0$
for all considered interaction strengths and in the whole range of electronic density.
Although the quantum correction decreases with $t'$ for the bcc lattice, 
as seen from the decreasing RPA exchange contribution (Fig. 3),
it still remains larger than the O(1) contribution even for $t'=0.5$ except in an
extremely narrow density range around $n=0.5$, 
where spin stiffness was found to be vanishingly small. 
However, the ferromagnetic state may be stabilized at larger $U$ 
as actually found for the fcc lattice with $t'=0.25$. The absence of ferromagnetic state on bipartite lattices in the considered range of intermediate coupling strength is in agreement with the existing results obtained using variety of 
techniques.\cite{Shastry, Vollhardt-jltp, Nolting-solid, Obermeier-largeU}
Also, the absence of ferromagnetism on fcc lattice for $t'=0$ is in agreement with recent DMFT and FLEX studies.\cite{Ulmke,Arita} 
%Although, ferromagnetic state has been reported for $t'=0$ in a wide range of densities at %significantly large interaction strength within MRPA,\cite{Rumch} 
%it remains to be seen how this result is modified when the frequency and momentum dependence %of vertex corrections are also incorporated.

\begin{figure}
\begin{center}
\vspace*{-18mm}
\hspace*{-12mm}
\psfig{figure=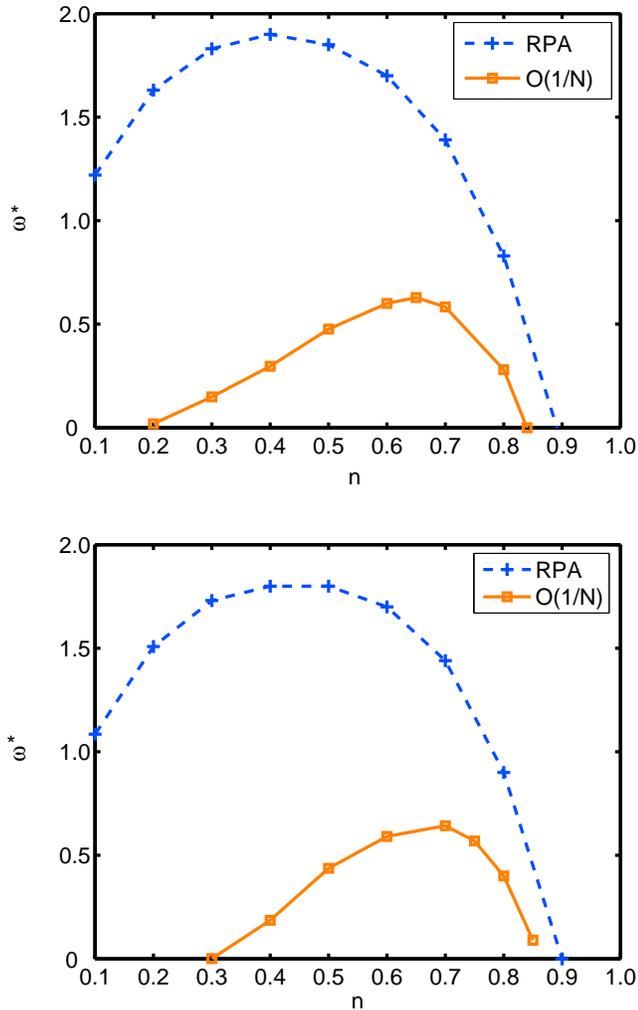,width=118mm}
\vspace*{-15mm}
\end{center}
\caption{The dominant-mode energy, 
corresponding to modes with large momenta near zone boundary,
suffers a strong suppression at low densities due to quantum corrections,
restricting stability to an intermediate range of densities,
shown for the fcc lattice at $U/W=1.0$ for $t'=0.25$ (upper) and 0.5 (lower).}
\end{figure}

Hereafter, the discussion refers to the fcc lattice specifically.  
Fig. 5 shows the substantial reduction in spin stiffness when quantum corrections are  included. Results obtained by evaluating quantum corrections separately 
with and without the contribution of Stoner excitations highlight the effect of Stoner excitations 
on the stability of the ferromagnetic state.
The dominant effect of Stoner excitations arises from the fourth term in (9).
While the collective spin-wave excitations significantly reduce the spin stiffness from its RPA value, 
the contribution of Stoner excitations, 
which is particularly important in the low-density regime, 
considerably narrows the stable range of densities for $t'=0.25$.
Indeed, for $t'=0$, it is the Stoner contribution which destabilizes the ferromagnetic state in the entire density range.

For $t'=0.25$, the ferromagnetic state is unstable for weak coupling ($U/W=0.5$), 
but becomes stable in a limited density range for intermediate coupling 
($U/W$=1.0 and 1.5), and shows (Fig. 6) an optimization near $n=0.5$. These results are qualitatively similar to the DMFT results,\cite{Ulmke} 
where the ferromagnetic state was found to be stable for $U\geq W$ in a relatively wider density range with an optimal density $n=0.65$. However, spatial fluctuations have been speculated to further reduce the stable density range.\cite{Vollhardt-cond}  
For $t'=0.5$, the role of both spin-wave and Stoner excitations becomes less significant (Fig. 5), which follows from the behavior of the classical exchange contribution (Fig. 4),
resulting in stabilization of the ferromagnetic state even for $U/W=0.5$,
and significant enhancement in spin stiffness (Fig. 6) with decreasing $n$.

For both $t'=0.25$ and 0.5, 
increasing $U$ results in enhancement of spin stiffness as well as 
of the stable range of densities (Fig. 6). 
This is due to both enhanced RPA stiffness due to reduced exchange contribution and also reduced Stoner contribution.

\section{O($1/\cal N$) Quantum corrections for large-momentum modes}
We now turn to large-$\bf q$ (short-wavelength) modes and examine quantum corrections 
to magnon energies for ${\bf q}$ near the zone boundary. 
As the magnon DOS is strongly peaked near the upper end of the spectrum, 
and therefore magnon-mode summations will be dominated, 
especially in high dimensions, 
by contribution of modes with energy near the peak energy $\omega^*$,
we specifically consider "dominant modes" 
with RPA-level energy $\omega_{\bf q}^0 \approx \omega^*$. 
We have checked for several such ${\bf q}$ 
and find that the quantum correction is quantitatively very similar. 
The behavior of the corrected magnon energy $\omega_{\bf q}$ with electron filling $n$ 
is of interest as this energy scale provides a quantitative measure of $T_c$ within the spin-fluctuation theory, thus allowing for a quantitative comparison with the $T_c$ vs. $n$ behavior within the DMFT which incorporates only local fluctuations. 

For finite ${\bf q}$, the magnon energy $\omega_{\bf q}$ is obtained from the pole condition $1-U\ Re \phi({\bf q},-\omega_{\bf q}) = 0$, 
where $\phi({\bf q},\omega) = \phi^{(0)}({\bf q},\omega) + \phi^{(1)}({\bf q},\omega)$. 
The four diagrammatic contributions to the first-order quantum correction
$\phi^{(1)}({\bf q},\omega)$ for arbitrary ${\bf q},\omega$ have been 
given earlier.\cite{As-goldstone}
While the bare particle-hole propagator $\phi^{(0)}({\bf q},\omega)$ remains real in 
the relevant $\omega$ range, the quantum correction $\phi^{(1)}({\bf q},\omega)$ is complex for any finite $\omega < 0$ due to the coupling with charge fluctuations. Evaluation of the $\Omega$ integral is similar as described below (9), and both collective and Stoner excitations are included. 

Fig. 7 shows the behavior of the dominant-mode energy $\omega_{\bf q}$ 
with electron filling for $t'=0.25$ and 0.5.
The nearly similar behavior shows the relatively much weaker sensitivity 
of large-momentum modes on details of band dispersion as compared to small-$q$ modes. 
The strong suppresion at low densities, a trend already present at the RPA level, 
results in an intermediate-density range of stability. 
This behavior of the dominant-mode energy 
is in excellent quantitative agreement with that of $T_c$ 
obtained in the DMFT study,\cite{Ulmke}
both with respect to the optimal density ($n=0.65$) 
as well as the stable range of densities ($0.20 < n < 0.85$). 
Furthermore, for a realistic bandwidth $W \sim 5$ eV for 3d transition metals, e.g. Ni,
the optimal magnon energy translates to approximately 0.15 eV,
which will be further reduced due to thermal magnon excitation at finite temperature,
bringing the energy scale also in good agreement 
with the optimal $T_c$ value of $\sim 0.05$ within DMFT. 

The substantial decrease in magnon energy at large momenta (Fig. 7)
has been referred to as 'anomalous softening',\cite{Hwang} 
and has been observed experimentally in ferromagnetic manganites.
Anomalous softening has also been investigated theoretically 
within an orbitally-degenerate Hubbard model\cite{Khaliullin} and the ferromagnetic Kondo lattice model.\cite{Nolting-FKLM} 

Comparison of Figs. 5 and 7 shows a marked contrast in the behavior of magnon energy 
for long- and short-wavelength modes for $t'=0.5$, 
whereas the behavior is qualitatively similar for $t'=0.25$. 
For $t'=0.5$, while positive spin stiffness in the low-density range 
indicates stability with respect to long-wavelength fluctuations, 
the vanishing of dominant-mode energy $\omega^*$ for short-wavelength modes
signals the spontaneous onset of strong local fluctuations.
On the other hand, while $\omega^*$ shows a peak near  $n \approx 0.7$, 
indicating stability with respect to local fluctuations,
the vanishing spin stiffness near $n \approx 0.7$
indicates loss of long-range magnetic order. 
That the stability of the ferromagnetic state with respect to 
long- and short-wavelength fluctuations can differ so substantially is a significant feature of our approach. 
The above results show that the loss of magnetic order and $T_c$ are determined by different mechanisms 
in the two density regimes ---
onset of strong local fluctuations in the low-density range and long-range fluctuations in the high-density range. 

%\begin{figure}
%\begin{center}
%\vspace*{-75mm}
%\hspace*{-30mm}
%\psfig{figure=fig8.ps,width=140mm}
%\vspace*{-80mm}
%\end{center}
%\caption{Variation of dominant-mode energy with interaction strength at a fixed density ($n=0.5$). In this range of U, %energy of long-wavelength magnons becomes negative at O$(1/{\cal N})$.}
%\end{figure}

\section{conclusions} 
Lattice dependence of correlation effects in a band ferromagnet
was investigated within a systematic inverse-degeneracy ($1/{\cal N}$) expansion 
which allows for self-energy and vertex corrections in the magnon propagator
to be incorporated in a spin-rotationally-symmetric scheme, 
so that the Goldstone mode is explicitly preserved. 
The resulting theory is therefore in accordance with Mermin-Wagner theorem
and hence allows for long-wavelength, low-temperature, and low-dimensional studies,
thus providing a complementary approach to the dynamical mean field theory. 

The first-order (O($1/{\cal N}$)) quantum corrections to magnon energies, 
evaluated separately with and without the contribution of Stoner excitations,
were investigated for different 
lattices (sc, bcc, fcc), 
band dispersions ($t'=0, 0.25, 0.5)$, 
interaction strengths ($U/W=0.5, 1.0, 1.5$),
and wave-vectors ($q$).
This provided a comprehensive picture of the 
interplay of lattice, dispersion, and interaction effects
on the stability of the ferromagnetic state 
with respect to both long- and short-wavelength fluctuation modes.

The present investigation supports the idea that ferromagnetism is a 
generic feature of the Hubbard model in the intermediate and strong-coupling limit 
provided the DOS is sufficiently 
asymmetric and strongly peaked near a band edge (fcc lattice with $t'$).
Quantum corrections are strong enough to destabilize the ferromagnetic state
for symmetric-DOS systems (bipartite lattices such as sc and bcc with $t'=0$),
as well as for weakly-asymmetric-DOS systems (fcc lattice with $t'=0$) 
in the whole range of densities and for all considered $U$ values,
in broad agreement with DMFT results. 
For the bcc lattice, we find instability of the ferromagnetic state 
even for finite $t'$ (in the range $t'\lesssim 0.5$).  

Due to the similarity between the correlation-induced (quantum) 
and the RPA (classical) exchange contributions to spin stiffness, 
a quantitative understanding of the quantum corrections,
and thus of the ferromagnetic-state stability with respect to long-wavelength fluctuations 
could actually be obtained from the RPA exchange contribution 
and its competition with the delocalization (hopping) contribution.

While the hopping contribution 
(for which the quantum correction identically vanished due to an exact cancellation) 
was found to vanish as $n\rightarrow 1$ and increase monotonically as $n\rightarrow 0$, 
both essentially lattice independent,
the exchange contribution was found to be highly sensitive to band dispersion and 
distribution of spectral weight in DOS, and therefore strongly lattice dependent.

For the fcc lattice with $t'=0.25$, the ferromagnetic state was found to be stable
in an intermediate range of densities near $n=0.5$, with both spin stiffness and 
stable range of densities increasing with interaction strength. 
Also, the density range for stability against long-wavelength fluctuations 
was found to be significantly reduced as compared to that against short-wavelength fluctuations, in agreement with the general expectation when spatial fluctuations
are included.\cite{Vollhardt-cond} 

For $t'=0.5$, while stiffness and stable range of densities similarly increase with $U$,
the enhancement in spin stiffness with decreasing $n$ is in sharp contrast to the 
low-density behavior of the $t'=0.25$ case, which demonstrated the subtle dependence of quantum corrections on details of the band dispersion. 
For short-wavelength modes near zone boundary, however,  
the behavior of the dominant-mode energy (corresponding to magnon-DOS peak),
was found to be quite similar for the two cases $t'=0.25$ and 0.5,
with an optimization near $n=0.65$.
The behavior of the dominant-mode energy,
which provides a quantitative measure of $T_c$ within the spin-fluctuation theory 
for high-dimensional systems, 
was also found to be qualitatively very similar to the 
$T_c$ vs. $n$ behavior within DMFT which incorporates only local fluctuations,
suggesting that the physically important quantum corrections are already 
incorporated at the first-order level. 

The case of fcc lattice with $t'=0.5$ was found to exhibit contrasting behavior 
with respect to long- and short-wavelength fluctuations.
The low-density regime is characterized by 
large stiffness but vanishing dominant-mode energy $\omega^*$, 
indicating stability against long-wavelength fluctuations but loss of magnetic order due to 
strong local fluctuations. On the other hand, 
the high-density regime is characterized by positive $\omega^*$ but vanishing stiffness,
indicating locally stable ferromagnetic order
but spontaneous onset of long-wavelength fluctuations.
The instability of the collinear ferromagnetic state 
in the low- and high-density regimes therefore involves quite different mechanisms. 

Thermal excitation of magnons and Stoner excitations at finite temperature 
result in magnetization correction due to spectral-weight transfer across the Fermi energy.
Finite-temperature magnon renormalization due to corresponding additional processes 
in the particle-hole propagator $\phi({\bf q},\omega)$,
which will yield the thermal decay of spin stiffness, magnon energy, and magnetization,
and thereby determine $T_c$, is currently under investigation.

%The extension of these investigations to the multi-orbital Hubbard model, 
%where correlation effects are expected to be relatively weak, including various other %lattices is a part of future study. 

\section{Acknowledgments}
One of us (SP) gratefully acknowledges financial support from CSIR.

\end{document}